\documentclass[aps,prl,twocolumn,groupedaddress,showpacs,tighten]{revtex4}
\usepackage{graphicx}
\usepackage{times}
\def\beq{\begin{equation}}
\def\eeq{\end{equation}}
\def\bey{\begin{eqnarray}}
\def\eey{\end{eqnarray}}

\def\lsim{\mathrel{\raise.3ex\hbox{$<$\kern-.75em\lower1ex\hbox{$\sim$}}}}
\def\gsim{\mathrel{\raise.3ex\hbox{$>$\kern-.75em\lower1ex\hbox{$\sim$}}}}

\begin{document}

\title{Constraints on Light Dark Matter From Core-Collapse Supernovae}
\author{Pierre Fayet$^{1}$, Dan Hooper$^{2}$ and G\"unter Sigl$^{3}$}

\affiliation{$^1$Laboratoire de Physique Th\'eorique de l'ENS, UMR 8549 CNRS, 24 rue Lhomond, 75231 Paris Cedex 05, France
 \\$^2$Fermi National Accelerator Laboratory, Particle Astrophysics Center, Batavia, 
IL 60510-0500,
 USA \\$^3$GReCO, Institut d'Astrophysique de Paris, CNRS, 98 bis boulevard Arago, 75014 Paris,
 France; \\
 AstroParticule et Cosmologie (APC), 11, place Marcelin Berthelot, 75005 Paris, France}

\date{February 18, 2006}

\begin{abstract}
We show that light \,($\simeq$ 1 -- 30 MeV) dark matter particles can play a significant role in core-collapse supernovae, if they have relatively large
annihilation and scattering cross sections, as compared to neutrinos.
We find that if such particles are lighter than $\,\simeq$ 10 MeV and reproduce 
the observed dark matter relic density, supernovae would cool on a much
longer time scale and would emit neutrinos with significantly smaller energies than in the
standard scenario, in disagreement with observations.
This constraint may be avoided, however, in certain situations for which the neutrino--dark matter scattering cross sections remain comparatively small.
\end{abstract}
\pacs{95.35.+d; 97.60.Bw
\hspace{1cm} FERMILAB-PUB-06-001-A \hspace{.5cm} LPTENS-06/04}
\maketitle

The identity of our Universe's dark matter is one of the most interesting questions in modern cosmology. Although a wide range of viable particle candidates have been proposed, none has been confirmed experimentally. The dark matter candidates most often studied are
weakly-interacting particles with masses in the $\,\sim\,$ 100 GeV to TeV scale;
neutralinos in supersymmetric theories being one prominent example.

Such dark matter particles should not be too light,
otherwise they could not annihilate sufficiently.
Still it is possible to consider light dark matter (LDM) particles,
with the right relic abundance to constitute the non-baryonic dark matter of the Universe,
provided one also introduces new efficient mechanisms responsible for their annihilations.
Such annihilations into, most notably, $e^+e^-\!$,
could correspond to the exchanges of new heavy (e.g. mirror) fermions
(in the case of light \hbox{spin-0} dark matter particles), or 
of a new neutral gauge boson $U$ \cite{boehmfayet}, light but very weakly coupled \cite{fayet:1980rr}, and still leading to relatively ``large'' annihilation cross sections.

The subsequent observation by the INTEGRAL/SPI experiment of a bright 511 keV
$\gamma-$ray line from the galactic bulge \cite{integral}
could then be viewed as a sign of the annihilations of positrons
originating from such light dark matter particle annihilations \cite{boehm511,Boehm:2003ha}.
These LDM particles, explaining both the {\it \,non-baryonic dark matter\,} of the Universe
and {\it \,the 511 keV line}, may have spin-$\frac{1}{2}$ as well as spin-0 \cite{fermion}.
They could even potentially improve the agreement
between the predicted and observed abundances of primordial $^2$H and $^4$He,
as long as LDM particles are not too strongly coupled to neutrinos \cite{bbn}.

Although astrophysical sources such as hypernovae have also been proposed as the source of the required positrons \cite{hypernovae}, 
such an origin seems in contradiction with the large extent of the 511 keV emission zone
\cite{jeanetal} (even if the uncertainties on the low-energy positron propagation and hypernova rate in the bulge are such that it is premature to conclude \cite{parizotetal}).
We shall therefore focus on the light dark matter interpretation of the 511 keV line.
(Other exotic particle physics scenarios which could generate this emission
have been proposed in \cite{otherexotic}.)

Given the large rate of positrons produced, smaller dark matter masses tend to be preferred,
to avoid an excessive production of unobserved $\gamma-$rays~\cite{fermion}.
More specifically, if such particles are heavier than 20--30 MeV,
internal bremsstrahlung (and bremsstrahlung) photons are likely to exceed the observed number of $\gamma-$rays from the galactic bulge \cite{beacom}.
And, if they were heavier than even 3 MeV, the $\gamma-$rays generated through the resulting $e^+e^-$ annihilations might also be inconsistent with observations \cite{beacom2}.

If MeV-scale dark matter particles do exist, they will be thermally generated in the core of collapsing stars. The presence of these particles can affect
thermal freeze out of weakly-interacting neutrinos, depending on their mass, and
annihilation and elastic scattering cross sections.

\begin{figure}

\includegraphics[width=0.45\textwidth,clip=true]{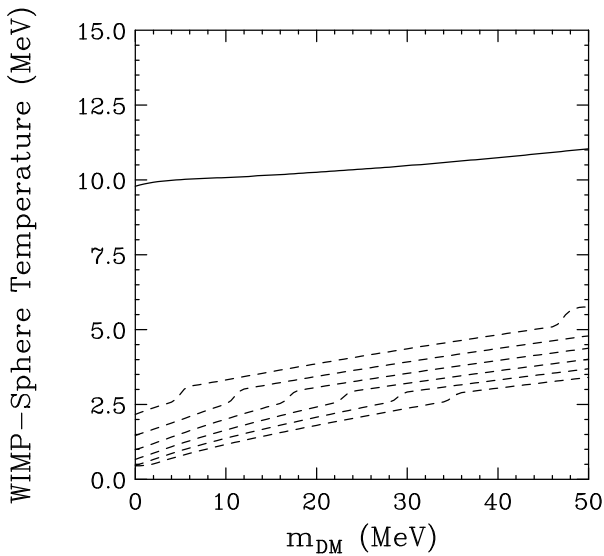}

\caption{Temperature of the LDM-sphere $\,T_{DMS}\,$ for $\,\sigma_{\rm ann}v_{\rm r}\propto v^2$
($P$-wave), normalized so as to lead to the observed relic density of dark matter.
The dashed lines represent various hypothesis for the LDM-nucleon elastic scattering cross section,
taken to be $\,\propto T^2$ and \,(from top to bottom) 1, 10, $10^2$, 10$^3$, 10$^4$, and 10$^5$ times larger than the corresponding neutrino-nucleon cross section.
The solid line shows the case of a massive $\nu_\tau$, \,for comparison.
\vspace{-5.5mm}
}
\label{tns}

\end{figure}

Ordinary neutrinos stay in thermal equilibrium through weak interactions 
down to temperatures $\simeq 2$
or 3 MeV during the expansion of the Universe, and down to $\simeq 8$  MeV or so, 
in supernovae explosions.
Light dark matter particles of mass $m_X\,$ annihilate into ordinary ones,
staying in equilibrium until they decouple. This occurs, during the expansion of the Universe,
at $\,T_F\!=m_X/x_F$, with $\,x_F\!\simeq 17$. In a supernova explosion LDM
particles will remain in chemical equilibrium with other particles, until the temperature drops down to some value
$\,T_{\rm DMS}$, \,to be determined later (cf. Fig.\,\ref{tns}).

As long as their abundance remains sufficient, these light dark matter particles can also influence the behavior of neutrinos in a supernova by having relatively ``large'' interactions with them, e.g. through $\,U$ exchanges \cite{boehmfayet}
\footnote{This appears as the counterpart of situations in which the propagation of neutrinos could affect 
the dark matter properties \cite{bfs}.}.
Neutrinos may then be kept longer in thermal equilibrium as a result of stronger-than-weak interactions with LDM particles, so that their decoupling temperature, in supernovae explosions, would be significantly lower than 
in the Standard Model,
if dark matter particles are sufficiently light. A crucial ingredient for this discussion will then be the magnitude 
of the neutrino--LDM \,elastic scattering cross section.

We now consider quantitatively these effects 
through a simple model based on the diffusion approximation, largely following Ref.~\cite{sigl}. We begin with the transport equation:
\begin{equation}
\label{dotn}
\dot{n} + \vec \bigtriangledown .\vec \phi \ \,=\,\ -\, \sigma_{\rm ann}v_{\rm r}\, (n^2 - n^2_{\rm{eq}})\ \ ,
\end{equation}
where $\vec\phi\,$ is the LDM flux, $n$ the number density of LDM particles and $n_{\rm{eq}}$
its equilibrium value for a LDM mass $m_X$ at temperature
$T$ with vanishing chemical potential.
$\,\sigma_{\rm ann}$ is the LDM, anti-LDM annihilation cross section (or self-annihi\-lation cross section
if the LDM is its own antiparticle\,\footnote{The factor $\frac{1}{2}\,$
that ought to be present in the r.h.s. of Eq.~(\ref{dotn})
then disappears since {\it two} self-conjugate LDM particles are removed in each annihilation.}), and $v_{\rm r}$ is representative of the relative velocity
of the two annihilating particles.

In the following, cross sections will be taken at typical thermal energies.
We now adopt the diffusion approximation, $\vec \phi=-D\,\vec \bigtriangledown n$, where
$D=\lambda v/3$ is the diffusion coefficient and
$\lambda=(\sum_i n_i\sigma_{Xi})^{-1}$ the LDM mean free path, which depends
on the densities $n_i$ of all the particle species (nucleons, electrons and positrons, $\nu$'s and $\bar \nu$'s, and in principle dark matter particles as well) with which the LDM interacts,
with cross sections $\,\sigma_{Xi}$. Assuming spherical symmetry and stationarity ($\dot n=0$), we can write the transport equation as:
\begin{equation}
D n^{''} + \bigg(D^{'} + \frac{2D}{r}\bigg)\, n^{'} \ =\ \,
\sigma_{\rm ann}v_{\rm r} \, (n^2 - n^2_{\rm{eq}})\ \,,\label{diff_wimp1}
\end{equation}
where the primes denote derivatives with respect to radius $r$.

We define the ``LDM-sphere" as the surface beyond which LDM annihilations 
(into $\,e^+e^-,\ \nu\bar \nu, \ ...$) "freeze out" such that the
LDM number is effectively conserved further out.
The number density of LDM particles inside of this LDM-sphere
should approach its equilibrium value $\,n_{\rm{eq}}$.
The radius of the LDM-sphere, $R_{\rm{DMS}}$, can be estimated by solving
\begin{equation}
\bigg| \,D n^{''}_{\rm{eq}} + \bigg(D^{'} + \frac{2D}{r}\bigg)\,n^{'}_{\rm{eq}}
\,\bigg|_{R_{\rm{DMS}}} =\ 
\bigg[\,\sigma_{\rm ann}v_{\rm r}\, n^2_{\rm{eq}}\,\bigg]_{R_{\rm{DMS}}}.\label{diff_wimp2}
\end{equation}

The radius of the surface of last scattering of LDM particles is found by solving
$\,\int_{R_{\rm{LS}}}^{\infty} dr\ \sum_i n_i(r)\,\sigma_{Xi} \simeq 1\,.$
%
%%%\begin{equation}
%%%\int_{R_{\rm{LS}}}^{\infty} dr\ \sum_i n_i(r)\  \sigma_{Xi}\, \simeq\, 1\ ,
%%%\label{scatt_wimp}
%\end{equation}
%
\,We shall concentrate here on LDM scatterings on nucleons (in practice mostly neutrons)
 with density $n_N(r)$,
and elastic cross section $\sigma_{XN}$, the actual $R_{LS}$ radius being at least as large as the one we shall estimate by disregarding the other species.
This conservative assumption is sufficient to demonstrate that LDMs
(and therefore eventually neutrinos, with which these LDM particles are normally coupled) decouple at lower densities and temperatures than
in the Standard Model \footnote{The scatterings 
of dark matter particles with $\,e^-\!,\ e^+\!, \ \nu$ and $\bar\nu$'s 
may also be important, in which case $R_{LS}$ would be larger than considered here
(solely from nucleons).
The same may occur for the scatterings of LDM particles with themselves,
as in many situations they are expected to interact more strongly with themselves than with other
(Standard Model) particles \cite{fermion}.}.

For the situations of interest to us, the LDM sphere lies within 
the last scattering sphere, so that the diffusion approximation 
is justified. Indeed, the annihilation cross sections of LDM particles 
are normally comparable to LDM scattering cross sections with ordinary particles\,\footnote{Except
of course in specific situations for which $\,\sigma_{\rm ann}\,$ gets reinforced as the exchanged particle
(e.g. a $U$ boson) can be nearly on-shell. $\sigma_{\rm ann}\,v_{\rm r}\,$ 
could then be large while scattering cross sections would be significantly smaller.}, so that if LDMs can still annihilate they can still also scatter.

To determine the LDM-sphere and surface of last scattering for a light dark matter particle, 
we must adopt a LDM mass and a set of (annihilation and scattering) cross sections, 
as well as a distribution of nucleons $n_N(r)$ and their temperature $T(r)$ in the proto-neutron star. 
For the latter, we will use the following parameterizations, which should be reasonable 
within the range $\sim$15--100 km we are concerned with  \cite{private}:

\pagebreak

\vspace*{-9mm}
\begin{equation}
\hbox{\small$\displaystyle
n_N(r) \simeq \left\{ \begin{array}{lll}
6 \times 10^{35} \, {\rm cm}^{-3} \, 
\bigg(\frac{23 \, {\rm km}}{r} \bigg)^{7.8}, \,\,\,\,\,\, r < 23 \, {\rm km}\,,  \\
6 \times 10^{35} \, {\rm cm}^{-3} \, 
\bigg(\frac{23 \, {\rm km}}{r} \bigg)^{12.8}, \,\,\,\,\,\, 23 \, {\rm km} < r < 44 \, \rm{km},  \\
1.2 \times 10^{33} \, {\rm cm}^{-3} \, 
\bigg(\frac{44 \, {\rm km}}{r} \bigg)^{3.0}, \,\,\,\,\,\, r > 44 \, {\rm km}\,, 
\end{array}
\right. \;
$}
\end{equation}

\vspace{-1mm}
\noindent
and

\vspace{-7mm}

\begin{equation}
T(r) = \left\{ \begin{array}{ll}
5.2 \, {\rm MeV} \, \bigg(\frac{25 \, {\rm km}}{r} \bigg)^{2.6}, \,\,\,\,\,\, r < 25 \, {\rm km}\,, \\
5.2 \, {\rm MeV} \, \bigg(\frac{25 \, {\rm km}}{r} \bigg)^{1.2}, \,\,\,\,\,\, r > 25 \, {\rm km}\,, \\
\end{array}
\right. \;
\end{equation}
These conditions are characteristic for the first few seconds over which
most of the cooling takes place, in the standard scenario.

We note that the presence of LDM could considerably modify profiles compared
to the standard scenario. However, we will find (cf.~\,Fig.\,\ref{tns})
that LDM reproducing the relic
density and having $\,\sigma_{XN}\sim\,\sigma_{\rm ann}$ are so strongly coupled
that they essentially stay in equilibrium as long as they are not Boltzmann
suppressed, so that they freeze out at temperatures $\,T_{\rm DMS}\lesssim m_X/3\,$
for LDM masses of interest here.

{\it Annihilation cross sections.}
The magnitude of the annihilation cross section of LDM particles into ordinary ones, 
at cosmological freeze out time, is fixed 
by the relic density requirement \cite{boehmfayet,fermion}. This leads to 
$\,(\sigma_{\rm ann}v_{\rm r}/c)_{\hbox{\scriptsize F}} \,\simeq \hbox{a few\ pb}$.

In the preferred case of a $P$-wave annihilation cross section 
(or at least $P$-wave dominated at freeze-out), 
$\,\sigma_{\rm ann}v_{\rm r}\,$ is roughly proportional to $\,v^2$, so that
$\,(\sigma_{\rm ann}v_{\rm r})_{P\hbox{-}{\rm wave}} \simeq (3\times10^{-35}\  {\rm cm}^{2}) \ \,v^2/c^2$. 
This also turns out to be the right order of magnitude 
for a correct injection rate of positrons in the galactic bulge. More precise statements about the respective roles of $P$-wave and $S$-wave contributions to LDM annihilations in the galactic bulge 
depend, of course, on more specific assumptions for the profile of the dark matter mass and velocity distribution 
adopted within the bulge \cite{ascrasera}.

We shall often have in mind a simple situation 
in which a scalar or fermionic dark matter particle annihilates (or interacts) through the virtual production (or exchange) \cite{boehmfayet,fermion} of a new light gauge boson, $U$ \cite{fayet:1980rr}, although the present analysis is more general.
This leads naturally to a $P$-wave annihilation cross section, both in the spin-0 case, 
and in the spin-$\frac{1}{2}$ case as well if the $U$ boson has vectorial (or mostly vectorial)
couplings to leptons and quarks. Axial couplings are already strongly 
constrained
from the non-observation of axionlike particles, and parity-violation effects in atomic physics~\cite{fermion,fayet:1980rr,pvat}.
For spin-0 particles there may also be $S$-wave contributions 
to the annihilation amplitudes, from the exchanges of new heavy (e.g. mirror) fermions.

{\it Results on the temperature of the LDM sphere.}
In Fig.~\ref{tns}, we plot the temperature of the LDM-sphere in the case of a
$P$-wave dominated annihilation cross section ($\propto v^2$), normalized to generate 
the measured dark matter relic density.
The results for a $S$-wave dominated
cross section are found to be very similar, since it has
the same value as a $P$-wave dominated one
(up to a factor $\simeq 2$) for a dark matter velocity
equal to its freeze-out value, $\,v_{\rm F}\simeq \,0.4 \ c$.
The dashed lines in Fig.~\ref{tns} correspond to various elastic scattering cross sections (see caption). The solid line shows, for comparison, the case of a weakly-interacting $\nu_\tau$ with a MeV-scale mass. While the
temperature $\,T\simeq 10$ MeV resulting for massless neutrinos in Fig.~\ref{tns} comes out a bit higher than the value $\simeq 8$ MeV from more detailed treatments~\cite{Keil:2002in},
what is most important is the relative value of the LDM and neutrino temperatures.

This shows that MeV scale LDMs will remain in equilibrium throughout the proto-neutron star at least down 
to relatively low temperatures $ T\simeq 3\,$ MeV,
{\it \,as an effect of the large values of the annihilation cross sections 
of LDM particles} (into ordinary ones). 
This occurs even if we do not assume rather high values of the scattering cross sections 
of LDM particles with ordinary ones.
Large scattering cross sections then contribute to further reinforce the effect
by increasing the LDM diffusion time allowing to keep LDM particles at chemical equilibrium down to even lower 
values of the temperature, possibly down to $\,T_{DMS}\simeq 1$ MeV, as illustrated by the lower dashed curves of Fig.~\ref{tns}.

{\it Consequences for the neutrino temperature.}
Thus light dark matter particles with relatively large annihilation cross sections
(as required from relic abundance) remain in
equilibrium down to lower temperatures, $\,T \lesssim 3$ MeV.
\,This feature may be transmitted to neutrinos,
that will themselves stay longer in thermal equilibrium as a result of their interactions with LDM particles,
{\it provided neutrino-LDM cross sections are also enhanced as compared to
ordinary neutrino cross sections}.

The kinetic equations  (\ref{diff_wimp1},\ref{diff_wimp2}) and the one fixing $R_{LS}$
%%%(\ref{diff_wimp1}-\ref{scatt_wimp})
are formally the same for neutrinos, substituting the
relevant cross sections for scattering and annihilation of neutrinos and the
equilibrium density of the relevant neutrino flavor. Inside the LDM sphere
the relevant quantities for neutrinos may be approximated as
\begin{eqnarray}
  D_\nu&\simeq&\frac{v}{3 \left(n_N\sigma_{\nu N}+n_{\rm eq,X}\,
  \sigma_{\nu X}\right)}\ \,, \nonumber\\
  \sigma_{{\rm ann}\,\nu}&\simeq&\sigma_{\,\nu\bar\nu\to \,X\bar X\,\,({\rm or}\,XX)}
  +\sigma_{\rm SM}\ ,
  \label{nu_quantities}
\end{eqnarray}
where cross sections are for the processes indicated as subscript (taking also into account the anti-LDM contribution if LDM particles are not self-conjugate), and
$\sigma_{\rm SM}$ indicates the Standard Model contribution. Note that
the LDMs are kinematically accessible by neutrinos for temperatures not
much lower than $m_X/3$.
\underline {If} indeed the cross sections for neutrino-LDM scattering and neutrino annihilation into LDMs
are comparable to the ones for LDM-nucleon scattering and LDM annihilations into leptons
(supposed to be ``large''),  respectively, 
the quantities in Eq.~(\ref{nu_quantities}) will
be dominated by the non-standard contributions.
This is because LDM cross sections are normally $\simeq $ a few pb (at freeze-out velocity, to give the appropriate relic abundance),
more than $\,\simeq 10^4$ larger than weak-interaction cross sections ($\,\simeq \,G_F^2\,T^2\,$), at the relevant energies.

This implies that neutrinos (if indeed they have relatively ``large'' interactions with LDM particles) 
should stay in chemical equilibrium {\it at least as long as the
LDMs do \,and $\,T\gtrsim m_X/3$}. We then conclude that $m_X\lesssim 10\,$MeV
would give rise to neutrino decoupling temperatures $\,\lesssim 3.3$ MeV for {\it all\,} flavors,
as compared to $\,\simeq 8$ MeV for $\nu_\mu$ and $\nu_\tau$ in the standard scenario.

This would make it quite unlikely to observe neutrinos with energy of order
$30-40\,$MeV, as have been observed from SN1987A~\cite{raffelt}, especially
for emission spectra that are suppressed at the highest energies compared to thermal distributions because the cross sections increase with energy~\cite{Keil:2002in}, 
in which case we can conclude that \textit {\,lighter LDM masses $\,\lsim 10$ MeV are
practically excluded.}

All this relies, of course, on the potentially ``large'' size of the neutrino-LDM scattering
and $\nu\bar\nu\to$ LDM's annihilation cross sections, normally expected to be comparable to the ``large'' LDM's  $\,\to\,e^+e^-$ annihilation cross section.

It is worth noting, however, that there are special situations for which
the $U$ boson would have no coupling at all (or suppressed couplings) to neutrinos \cite{U}. 
And that for a spin-0 LDM particle interacting with ordinary ones through the exchanges 
of heavy (e.g. mirror) fermions, the $\nu$-LDM interactions would be severely suppressed (as compared to  
electron or nucleon-LDM interactions), as a result of the chiral character of the neutrino field 
\cite{boehmfayet,fermion}. In both cases we end up with {\it no significant enhancement} 
of neutrino-LDM interactions, so that the presence of the LDM particles has no direct significant effect on 
the behavior of neutrinos, then still expected to decouple at $\,\simeq 8$ MeV (for $\nu_\mu$ and $\nu_\tau$), as usual. In such a case, \textit {no new constraint is obtained on the
mass $m_X$ of LDM particles.}

Furthermore, the above results may also be obtained, or understood, as follows.
Let us return to LDM particles rather ``strongly'' coupled to neutrinos (and nucleons), both types of particles decoupling at $T \lsim 3.3 $ MeV.
As  LDMs can then contribute, at most, as much to the
cooling flux as the neutrinos 
(due to fewer degrees of freedom), the
cooling time scale would be larger  than in the standard scenario by a factor $\gtrsim(8/3.3)^4/2\simeq20$ because the thermal flux is also $\propto T^4$. As SN1987A
observations were consistent with the standard cooling time scale of 10-20 s, 
such non-standard scenarios are then very strongly disfavored, to say the least.

The cooling time scale can also be estimated by the diffusion time
$\,\tau_{\rm diff}\sim R_{\rm NS}^{\,2}/\lambda\,$. This is dominated by the
innermost regions
of the hot neutron star of size $R_{\rm NS}\simeq 10$ km, whose density
is not significantly modified by the presence of LDM. At a typical
temperature $T\simeq 30\,$MeV, 
$\sigma_{\nu N}\simeq11\,G_{\rm F}^2T^2/\pi\simeq 1.7\times 10^{-40}$ cm$^2$,
and at nuclear densities $n_{\rm eq}\sim n_N/100$. Thus,
for neutrino-LDM cross sections comparable 
to electron-LDM ones (i.e.~\,typically
$\,\gtrsim 4 \times10^{-36}\,{\rm cm}^2$,
\,so that $\,\sigma_{\nu X}\gtrsim \,10^4 \,\sigma_{\nu N}$),
\,the neutrino mean free path is dominated by interactions with
the thermal population of LDMs, so that $\,\lambda_\nu\sim(n_{\rm eq}\sigma_{\nu X})^{-1}
\lsim 0.3$ cm, as compared to $\lambda_\nu\sim(n_N\sigma_{\nu N})^{-1}
\sim 35$ cm in the standard scenario. The LDM mean free path is even
shorter, $\lambda_{\rm LDM}\sim(n_N\sigma_{XN})^{-1}\lsim 1.5\times10^{-3}$ cm, assuming $\,\sigma_{XN}\sim\sigma_{\nu X}\sim\sigma_{e X}$
(or even less if LDM self-interactions were to contribute significantly).
In the interior of the proto-neutron star
the energy flux is thus dominated by neutrinos.
The cooling time scale is a factor $\gtrsim 100$
larger than in the standard scenario, consistent with the previous argument.
This cooling time argument may be extended up to higher LDM masses
$\simeq 20$ or even 30 MeV, \,i.e. as long as LDMs are significantly present 
at $\,T \simeq 30$ MeV, and rather ``strongly'' coupled to neutrinos.

Given that about $3\times10^{53}\,$erg of binding energy has to be liberated
during $\tau_{\rm diff}$, in the relativistic regime the freeze out temperature
will scale as $T_\nu\propto \tau_{\rm diff}^{-1/4}$.
For $\sigma_{XN}\gtrsim 10^{4}\ \sigma_{\nu N}\,(T\sim30\,{\rm MeV})$,
this argument suggests $\,T_\nu$ will be a factor $\,\gtrsim 3\,$
times smaller than usual, as found previously.

In conclusion, we have demonstrated that light dark matter models with generically ``large''
cross sections fixed by requiring them to reproduce the relic dark matter density
are considerably disfavored by the resulting modification of core-collapse
supernova cooling dynamics if the dark matter mass is 
$\ \lesssim\,$ 10 MeV, at least.

Depending on how strict $\gamma-$ray constraints from the galactic bulge are,
the new supernovae constraint presented here could strongly disfavor 
the possibility that annihilating dark matter particles 
be the source of the 511 keV emission 
from the galactic bulge.  

Or, conversely, these new results could indicate that neutrino-LDM interactions
should {\it \,not} be enhanced, favoring a $\,U$ boson with no (or small) couplings to neutrinos
and/or a spin-0 dark matter particle interacting through heavy fermion exchanges.

We would like to thank Gianfranco Bertone, Thomas Janka and Georg Raffelt
for helpful discussions. 
DH is supported by the US Dep. of Energy and by NASA grant NAG5-10842.

\vspace{-9.5mm}
\phantom {a}

\end{document}